\documentstyle[aps,amsfonts,epsfig]{revtex}   

\newcommand{\beq}{\begin{eqnarray}}
\newcommand{\eeq}{\end{eqnarray}}

\newcommand{\pardis}{\langle \mu \rangle}

\newcommand{\proj}{\vec{\Phi}(x)}
\newcommand{\vproj}{\hat{\Phi}(x)}
\newcommand{\vpronc}{\hat{\Phi}}

\newcommand{\ie}{{\it i.e.\ }}
\newcommand{\eg}{{\it e.g.\ }}

\begin{document}     

\twocolumn[\hsize\textwidth\columnwidth\hsize\csname
@twocolumnfalse\endcsname
  
\draft

\title{Color confinement and dual superconductivity of the vacuum. III.}
 
\author{J.M. Carmona$^{a,1}$, M. D'Elia$^{b,2}$, 
A. Di Giacomo$^{c,d,3}$, B. Lucini$^{e,4}$, G. Paffuti$^{c,d,3}$}

\address{$^a$ Departamento de F\'{\i}sica Te\'orica, Universidad de Zaragoza, 
50007 Zaragoza, Spain}
\address{$^b$  Dipartimento di Fisica dell'Universit\`a di Genova and INFN,
Sezione di Genova, Via Dodecaneso 33, I-16146 Genova, Italy}
 \address{$^c$ Dipartimento di Fisica dell'Universit\`a, Via Buonarroti
2 Ed. B, I-56127 Pisa, Italy}
\address{$^d$ INFN sezione di Pisa, Via Vecchia Livornese 1291, I-56010
S. Piero a Grado (Pi), Italy}    
\address{$^e$ Theoretical Physics, University of Oxford,
1 Keble Road, OX1 3NP Oxford, UK}
\address{$^1$ e-mail address: jcarmona@posta.unizar.es}
\address{$^2$ e-mail address: delia@ge.infn.it}
\address{$^3$ e-mail address: digiaco, paffuti@df.unipi.it}
\address{$^4$ e-mail address: lucini@thphys.ox.ac.uk}
        
\maketitle

\begin{abstract}
It is demonstrated that monopole condensation in the confined phase
of $SU(2)$ and $SU(3)$ gauge theories is independent of the specific Abelian 
projection used to define the monopoles. Hence the dual excitations
which condense  in the vacuum to produce confinement must have magnetic
$U(1)$ charge in all the Abelian projections.
Some physical implications of this result are discussed.
\end{abstract}
]

\section{Introduction}              

This paper is the third of a series, in which we study the dual
superconductivity of the QCD vacuum \cite{'tHooft75,'tHooft81,Mandelstam76} as a 
mechanism for confinement of color.
In the first two papers \cite{artsu2,artsu3} we have detected 
condensation of monopoles in the confining phase by means of a disorder
parameter $\pardis$, which is the vacuum expectation value of 
a magnetically charged operator, $\mu$. $\pardis$ was determined by simulating
the theory on a lattice: it is non zero in the confining phase, and tends to 
zero  at the deconfining transition, above which it vanishes.

The connection of $\pardis$ to confinement was proved by a quantitative
determination of the critical indices and of the critical coupling.
In \cite{artsu2}, $SU(2)$ gauge theory was studied, while \cite{artsu3}
was devoted to $SU(3)$ with similar results.

Magnetic charges in gauge theories are defined by a procedure known as Abelian 
projection~\cite{'tHooft81}: to every local field $\Phi$ belonging to the adjoint representation
$(N_c -1)$ $U(1)$ fields can be associated, and to each of them a conserved 
magnetic charge. In fact there exists a functional infinity of monopole 
species, $N_c -1$ for each field $\Phi$, which in principle can condense in 
the vacuum and confine the corresponding $U(1)$ electric charge by the dual
Meissner effect. It is not known a priori if monopole condensations in different
Abelian projections are independent phenomena.

The indication obtained in \cite{artsu2,artsu3} by analysis of a number
of different choices of $\Phi$ was that all of them show the same behaviour, 
so that they are equivalent to each other.

The possibility that in some way all the Abelian projections could be physically
equivalent was first advocated in Ref. \cite{'tHooft81}.
In this paper we add strong evidence for that equivalence.

In order to explain what we do, let us first recall how magnetic charges are
associated to any field $\Phi$ in the adjoint representation. We shall do it
for $SU(2)$ to simplify notations; extension to $SU(N)$ only adds formal 
complications.

Let $\proj $ be a field in the adjoint representation (color vector),
and let $\vproj$ be its color orientation,
\beq
\label{projop}
\vproj  \equiv {\frac{\proj}{|\proj|}} \; .
\eeq
$\vproj$ is well defined, except at zeros of $\proj$.

Define a gauge invariant field strength $F_{\mu\nu}(x)$ \cite{'tHooft74}
\beq
\label{fabel}
F_{\mu \nu} = \vpronc \cdot \vec{G}_{\mu \nu} - \frac{1}{g}
\left( D_{\mu} \vpronc \wedge D_{\nu} \vpronc \right) \cdot \vpronc
\eeq
where $\vec{G}_{\mu \nu} = \partial_\mu \vec{A}_\nu - \partial_\nu \vec{A}_\mu
+ g \vec{A}_\mu \wedge  \vec{A}_\nu$ is the gauge field strength and
$D_\mu \hat{\Phi} = (\partial_\mu + g \vec{A}_\mu \wedge ) \hat{\Phi}$ is the
covariant derivative of $\hat{\Phi}$.

Both terms in the right-hand side of eq. (\ref{fabel}) are separately
gauge invariant and color singlets: their combination is chosen in such
a way that bilinear terms $A_\mu A_\nu$, $A_\mu \Phi$ and $A_\nu \Phi$ cancel.
Actually, by simple algebra,
\beq
\label{fabel2}
F_{\mu \nu} = 
\vpronc \cdot (\partial_\mu \vec{A}_\nu - \partial_\nu \vec{A}_\mu) 
- \frac{1}{g} \left( \partial_{\mu} \vpronc \wedge 
      \partial_{\nu} \vpronc \right) \cdot \vpronc \; .
\eeq

If we transform to a gauge in which $\vproj = $ constant in space-time,
the last term cancels and
\beq
\label{fabel3}
F_{\mu \nu} = 
 \partial_\mu (\hat{\Phi} \cdot \vec{A}_\nu) -  
 \partial_\nu (\hat{\Phi} \cdot \vec{A}_\mu) 
\eeq
is an Abelian field strength. Such a gauge transformation is called an 
Abelian projection. It is in general a singular transformation which exposes 
monopoles at the sites where $\proj = 0$.

If $F^{\star}_{\mu \nu} = \frac{1}{2} \epsilon _{\mu \nu \rho \sigma} F^{\rho
\sigma} $ is the dual field to $F_{\mu \nu}$, one can define a magnetic current
\beq
\label{mcurr}
j_{\mu} &=& {\partial}^{\nu} {F}^{\star}_{\mu \nu} \ ;
\eeq
$j_\mu$ is zero if Bianchi identities hold, but can be non zero in compact
formulations in terms of parallel transport, like lattice  formulation
\cite{Degrand}. In any case it follows from the antisymmetry of 
$F^{\star}_{\mu \nu}$ that 
\beq
\label{zerodiv}
\partial^{\mu} j_{\mu} = 0 \ .
\eeq 

In the dual superconductor view of color confinement the symmetry 
(\ref{zerodiv}) is expected to be realized  {\em \`a la Wigner} in the 
deconfined phase, and to be  broken {\em \`a la Higgs} in the  confined phase.
An operator $\mu$ which
carries magnetic charge can provide a disorder parameter to discriminate
between the two possibilities. Such an operator was developed and tested in
Refs \cite{ldd,Digiau1,Pieri,artsu2,artsu3}.

What was found in \cite{artsu2,artsu3} was that there is indeed dual 
superconductivity in a number of Abelian projections.
As explained in detail in Sect.~II, the full identification of the 
projected gauge requires to go to the gauge in which 
$\proj \cdot \vec{\lambda}$ is diagonal in color indices
($\vec{\lambda}$ are the generators in the fundamental representation),
with a fixed order of the eigenvalues.

One can diagonalize $\proj \cdot \vec{\lambda}$
up to the ordering of the eigenvalues, choosing it randomly, and still define an 
operator $\mu$ which creates a magnetic charge in that Abelian projection.
We show in this paper that the corresponding disorder parameter behaves 
exactly in the same way as the one with ordered eigenvalues.

We can even define a completely random Abelian projection, in which we do
not diagonalize any operator $\Phi$, but we take, e.g. for $SU(2)$,
$\hat{\Phi} = \sigma_3$, the nominal 3 axis used in the simulation, and 
define the corresponding $\mu$. Again we find that $\mu$ defined in this way 
behaves exactly in the same way as those defined in Refs. \cite{artsu2,artsu3}
and scales with the same critical indices.

The above Abelian projections are kind of an average over a continuous infinity
of Abelian projections, and the result demonstrates, beyond any doubt, 
the complete independence of dual superconductivity from the choice of 
the Abelian projection.

Our results are compatible with Ref. \cite{bari}, where our disorder
parameter in the random gauge
was computed by Schroedinger functional techniques.

There has been in the literature in the last years the idea that monopoles 
defined by a particular Abelian projection (the maximal Abelian projection)
are more relevant than others to confinement \cite{Suzuki,Stack}. We will 
discuss this issue in Sect.~V, where we draw conclusions from our results.

In Sect.~II the construction of the disorder parameter $\pardis$
will be recalled, to define the Abelian Projection with Random 
Ordering (APRO) and the Random Abelian Projection (RAP). 
In Sect.~III the numerical algorithms used will be discussed.
The results will be described in Sect.~IV. Sect.~V will close the paper 
with conclusions.

\section{Disorder parameter}
In this Section we recall the definition of the disorder parameter for confinement.

Let ${\cal O}$ be an operator which transforms in the adjoint representation of
the gauge group, \ie
\beq
{\cal O} =  \sum \lambda^a {\cal O}^a \, ,
\eeq 
with $\lambda^a$ the generators in the fundamental representation.
The Abelian projection technique \cite{'tHooft81} prescribes to fix the gauge 
by a gauge transformation in such a way that
\beq
\label{projdef}
{\cal O}_{gf} &=&  G^{\dagger} {\cal O} G = \mbox{diag}(o_1,...,o_N) \qquad \nonumber \\ 
& & \mbox{with}  \ o_1 < o_2 < ... < o_N \ .
\eeq
After Abelian projection, there is still a $U(1)^{N-1}$ gauge freedom left, 
since a transformation of the form
\beq
\Omega = \mbox{diag}(e^{i\omega_1},...,e^{i\omega_n}) \ , \qquad \sum \omega_i = 0
\eeq
does not change the gauge fixing condition, Eq. (\ref{projdef}).

After Abelian projection,
the gauge variables of $SU(N)$ are divided in two sets: the {\em photons} (the $N-1$ neutral fields
under the residual $U(1)^{N-1}$) and the {\em gluons} (charged fields with respect to the residual
symmetry). Abelian magnetic monopoles can arise at points where two 
eigenvalues of ${\cal O}$ are degenerate~\cite{'tHooft81}.
   
Condensation of Abelian monopoles defined by Abelian projection has been demonstrated numerically in
Ref.~\cite{artsu2} for $SU(2)$ and in Ref.~\cite{artsu3} for $SU(3)$. This has been done by constructing
an operator magnetically charged in a given Abelian projection
and by studying the behaviour of the vacuum expectation value of
that operator across the phase transition at finite temperature. In the language of
Statistical Mechanics we call that operator a {\em disorder operator} and its {\em vev} a {\em disorder parameter}, the terminology being that the 
weak coupling (deconfined) phase is the ordered phase. 
The construction can be done in different Abelian projections:
in Refs. \cite{artsu2,artsu3} a number of Abelian projections have been 
studied, and for all of them it was found that indeed
monopoles condense at low temperature, while the corresponding magnetic symmetry is implemented {\em \`a la Wigner}
at high temperature. Moreover, the disorder parameter scales with the correct critical indices
in the critical region and is independent of the choice 
of the Abelian projection. These results suggest that the 
observed behaviour of the disorder parameter
is generally independent of the Abelian projection and of the Abelian operator chosen.

Let us review the construction of the disorder parameter. We introduce a time-independent external field
\beq
\Phi_{i}(\vec{n},\vec{y}) = G e^{i T b_{i}(\vec{n}-\hat{i},\vec{y})} G^{\dagger} \ ,
\eeq  
where $G$ is the gauge transformation that diagonalizes the operator ${\cal O}$ according
to Eq.~(\ref{projdef}), $\vec{b}$ is the discretised transverse field (\ie $\vec{\nabla} \cdot \vec{b}$ =
0 on the continuum) generated at lattice spatial point $\vec{n}$ by a magnetic monopole sitting at
$\vec{y}$ and $T$ is a generator of the Cartan subalgebra.

Let $U_{\mu \nu}$ be the Wilson plaquette defined in the usual notations as
\beq
U_{\mu \nu}(x) = U_{\mu}(x) U_{\nu}(x+\hat{\mu}) 
& & ({U_{\mu}(x+\hat{\nu}}))^{\dag}
(U_{\nu}(x))^{\dag} \; ,
\eeq
where $x \equiv (\vec{n},t)$. We introduce a shift~$U_{i0}(\vec{n},0) \to \tilde{U}_{i0}(\vec{n},0)$
by inserting the external field $\Phi_{i}(\vec{n}+\hat{i},\vec{y})$
in the path ordered product at time zero, as follows:
\beq
\tilde{U}_{i0}(\vec{n},0) &=& U_{i}(\vec{n},0) 
\Phi_{i}(\vec{n}+\hat{i},\vec{y}) U_{0}(\vec{n}+\hat{i},0)\\  \nonumber
& &(U_{i}(\vec{n},1))^{\dag} (U_{0}(\vec{n},0))^{\dag} 
\eeq
and we define  $\tilde{U}_{\mu \nu}(x) \equiv U_{\mu \nu}(x)$ elsewhere.

The Wilson action for $SU(N)$ gauge theory is 
\beq
S = \beta \sum_{\mu \nu x} \left( 1 - \frac{1}{2N}
\left( U_{\mu \nu}(x) + \left(U_{\mu \nu}(x)\right)^{\dagger}\right) \right) \ ,
\eeq
where the sum extends over all the lattice points and directions.

By replacing in the previous equation the standard plaquette with the modified plaquette
$\tilde{U}_{\mu \nu}(i)$ we obtain the ``monopole'' action
\beq
S_M (\vec{y},0) = \beta \sum_{\mu \nu x} \left( 1 - \frac{1}{2N}
\left( \tilde{U}_{\mu \nu}(x) + (\tilde{U}_{\mu \nu}(x))^{\dagger}\right) \right) .
\eeq
The disorder parameter introduced in \cite{artsu2,artsu3} is given by
\beq
\langle \mu(\vec{y}_0,0) \rangle = 
\frac{\int \left( {\cal D} U \right) e^{-S_{M}(\vec{y}_0,0)}}
{\int \left( {\cal D}U \right)  e^{-S}} \ ,
\eeq
where the functional integral of $e^{-S}$ is taken with periodic boundary conditions and the integral
of $e^{-S_M}$ with $C^{\star}$-periodic boundary conditions \cite{Wiese1,Wiese2}, 
\beq
U_i(\vec{n},t=N_t) = U_i^* (\vec{n},t=0) \ ,
\eeq
$N_t$ being the temporal extension of the lattice and $U_i^*$ being the complex conjugated of $U_i$.

To study more in detail the dependence on the projecting operator, we will 
modify the definition of $\Phi$ as follows:
\begin{enumerate}
\item We choose a projecting operator and we diagonalise it but without fixing the order of the eigenvalues,
or better with the order of the eigenvalues randomly chosen:
\beq
\label{APRO}
\Phi_{i}(\vec{n},\vec{y}) = G P e^{i T b_{i}(\vec{n}-\hat{i},\vec{y})} P G^{\dagger} \ ,
\eeq
where $P$ is a random $N \times N$ permutation matrix. This corresponds to a sort of average of $\mu$ over  
the class of operators differing from ${\cal O}$ on each point by the order of the eigenvalues. We refer
to this case as Abelian Projection with Random Ordering (APRO).
\item We do not perform the Abelian projection, \ie we take
\beq
\label{RAP}
\Phi_{i}(\vec{n},\vec{y}) = e^{i T b_{i}(\vec{n}-\hat{i},\vec{y})} \ .
\eeq
This is equivalent to a sort of average of $\mu$ over all
the possible Abelian projections. We refer to this case as Random Abelian Projection (RAP).  
\end{enumerate}
As discussed in Ref. \cite{artsu2,artsu3}, a direct computation of $\pardis$ with Monte Carlo techniques is problematic, 
because this quantity has large fluctuations, being the exponential of
a sum over the physical volume.
A more convenient quantity to study in numerical simulations is \cite{ldd,artsu2,artsu3}
\beq
\rho = \frac {\partial}{\partial \beta} \log \pardis = \langle S \rangle_S -
\langle S_M \rangle_{S_M} \ .
\label{rhodef}
\eeq
$\rho$ is the difference of two average actions, the Wilson action and the modified action $S_M$ (the latter
being averaged with the modified measure $(({\cal D}U)e^{-S_M})/(\int({\cal D}U)e^{-S_M})$).
$\rho$ has smaller fluctuations and contains all relevant information.
The value of $\pardis$ is related to $\rho$ by the relationship
\beq
\pardis = \exp\left(\int_0^{\beta} \rho(\beta^{\prime})\mbox{d}\beta^{\prime}\right) \; .
\eeq

\section{Gauge fixing and simulation algorithms}
\label{algorithms}

We have determined the temperature dependence of $\rho$ for $SU(2)$
and $SU(3)$ pure Yang-Mills theories, for both definitions (\ref{APRO}) 
and (\ref{RAP}) of $\Phi$ on an asymmetric lattice $N_s^3 \times N_t$ with
$N_t \ll N_s$. 

For both definitions of $\Phi$, the simulation of the Wilson term,
$\langle S \rangle_S$, has been
performed on a lattice with periodic b.c., 
by using a standard mixture of heatbath and overrelaxed algorithms.

As for the APRO case, we have
chosen the Polyakov line as the operator to identify the Abelian projection, 
following the definition in Eq. (31) of \cite{artsu2} and 
Eq. (19) of \cite{artsu3},
with the only difference that at  each spatial point the ordering
of the eigenvalues is selected randomly among the possible different
permutations $n_p$ ($n_p = 2$ ($6$) for $SU(2)$ ($SU(3)$) pure gauge theory). 
This effectively corresponds to averaging over
$n_p^{N_s^3}$ different definitions of the Abelian projection.
The Abelian generator $F^8$ ($F^8 = \lambda^8/2$, with $\lambda^i$ the 
Gell-Mann matrices), has been chosen to define the monopole field
for the $SU(3)$ case. We use $C^{\star}$ boundary conditions in time
to compute $\langle S_M \rangle_{S_M}$ in Eq. (\ref{rhodef}).
In this case, as explained in Ref. \cite{artsu2}, it is not possible
to use a standard heatbath or overrelaxed algorithm to simulate
the modified action, since, \eg in the case of Polyakov projection,
the change of any temporal link induces a non linear change in the 
modified action. So we have performed simulations by using a 
 mixed (heatbath + overrelaxed) algorithm for the update of spatial links,
and a metropolis algorithm for the update of the temporal links.

In the case of the Random Abelian Projection instead, as it appears in 
Eq. (\ref{RAP}), one does not need any gauge fixing to define the monopole
field. As a consequence, the change of any link always induces a linear
change in the modified action. Therefore we have used a standard
(heatbath + overrelaxed) algorithm in this case.
In the $SU(3)$ case, the Abelian generator $F^8$ has been used again 
to define the monopole field.

\section{Numerical results}

The phase transition is known to be second order for $SU(2)$, weak first order
for $SU(3)$~\cite{Fuku}. As usual we shall speak of critical indices in both cases, meaning
for $SU(3)$ effective critical indices at small values of $(1 - T/T_c)$, but
not too small.

In Refs.~\cite{artsu2,artsu3} it was shown that the critical indices
of the 
confinement transitions for $SU(2)$ and $SU(3)$ did not depend on the particular 
type of Abelian projection used to define the monopole condensation. 
Here we will show numerically that the critical exponents, 
and also the value of the critical coupling,
are the same even for the Random Abelian Projection
and the Abelian Projection with Random Ordering,
both for $SU(2)$ and $SU(3)$

The critical behaviour of the disorder parameter $\pardis$ is
governed by an exponent $\delta$. For finite lattice sizes $(N_s^3\times N_t)$,
finite size scaling states that
\begin{equation}
\pardis = N_s ^{- \delta / \nu}
F \left(\frac{\xi}{N_s},\frac{a}{\xi},\frac{N_t}{N_s} \right) , 
\end{equation}
where $a$ and $\xi$ are respectively the lattice spacing
and the correlation length of the system.

Near the critical point, for $\beta < \beta_C$,
\begin{equation}
\xi \propto\left( \beta _C - \beta \right)^{- \nu} , 
\end{equation}
where $\nu$ is the
corresponding critical exponent. In the limit $N_s \gg N_t$  and
for $a / \xi \ll 1$, \ie sufficiently close to the critical point,
\begin{equation}
\label{scala1}
\pardis = N_s ^{- \delta / \nu}
\tilde{F} \left[N_s^{1/\nu}\left( \beta _C - \beta \right)\right]
\end{equation}
or equivalently
\begin{equation}
\label{scaling} 
\frac{\rho}{N_s^{1/\nu}} = f\left[ N_s^{1/\nu} 
\left( \beta _C - \beta \right)\right] .
\end{equation}
The ratio $\rho / N_s^{1/\nu}$ is a universal function of the
scaling variable 
\begin{equation}
x =   N_s^{1/\nu} \left( \beta _C - \beta\right) .
\end{equation}

We will use the known values of $\beta_C$ and $\nu$ of $SU(2)$
and $SU(3)$ pure gauge theories to see that scaling holds with the
present data. In order to obtain the critical exponent $\delta$, we
use an expression equivalent to Eq.~(\ref{scala1}),
\begin{equation}
\label{scala2}
\pardis = (\beta_c-\beta)^\delta {\cal F}(x) \; .
\end{equation}
From here we get
\begin{equation}
\frac{\rho}{N_s^{1/\nu}} = - \frac{\delta}{x} 
- \frac{{\cal F}'(x)}{{\cal F}(x)} \; .
\end{equation}
To obtain $\delta$ we need additional assumptions on the
unknown scaling function ${\cal F}(x)$. We will see that fits of good
quality are obtained with the simple parametrization
\begin{equation}
\label{eqfit}
\frac{\rho}{N_s^{1/\nu}} = - \frac{\delta}{x} - C \; ,
\end{equation}
where $C$ is a constant term. This form is suggested by the fact that 
when $x\to 0$, both ${\cal F}(x)$ and its derivative should go to a constant.

\subsection{$SU(2)$ gauge theory}

\subsubsection{The Random Abelian Projection}

\begin{figure}[tb]
\centerline{\epsfig{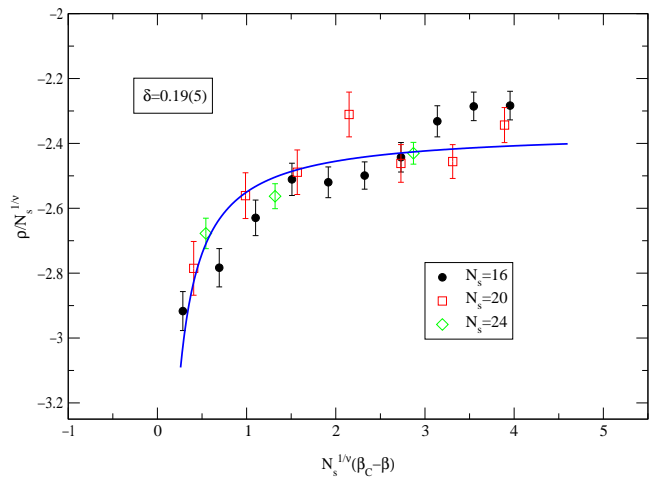}}
\vskip 0.5cm
\caption{Quality of scaling in the RAP case for $SU(2)$.}
\label{fig:su2_nogauge}
\end{figure}

The quality of the scaling, Eq.~(\ref{scaling}), for $SU(2)$ in the 
RAP case can be seen in Fig.~\ref{fig:su2_nogauge}. Here
we used the known values of $\beta_C=2.2986$ and $\nu=0.63$~\cite{karschsu2}.
The curve in the figure corresponds to the
best fit to Eq.~(\ref{eqfit}). We get $\delta=0.19(5)$, with
a $\chi^2/{\rm d.o.f.}\sim 1.5$, in good agreement with the value 
obtained in the plaquette and Polyakov gauges 
in Ref.~\cite{artsu2}, $\delta=0.20(8)$. 
\vskip 0.7cm

\begin{figure}[b]
\centerline{\epsfig{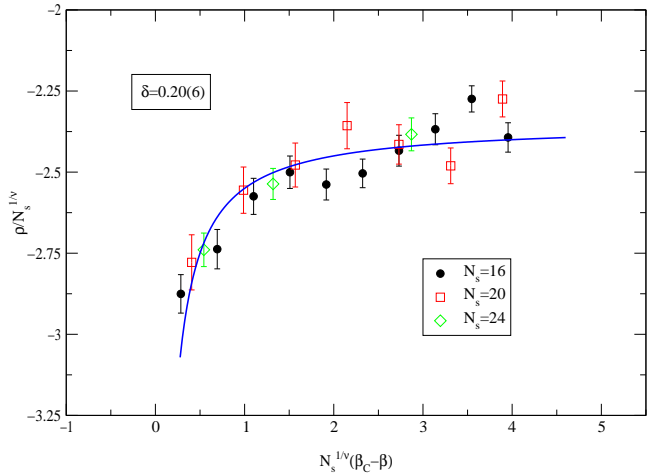}}
\vskip 0.5cm
\caption{Quality of scaling in the APRO case for $SU(2)$.}
\label{fig:su2_noorder}
\end{figure}     

\subsubsection{The Abelian Projection with Random Ordering}

Results obtained in the APRO case are shown in Fig.~\ref{fig:su2_noorder},
where again known values of $\beta_C$ and $\nu$ have been used.
The curve in the figure corresponds to the best fit to 
 Eq.~(\ref{eqfit}), which gives $\delta=0.20(6)$, with
a $\chi^2/{\rm d.o.f.}\sim 1.1$. The agreement with the results
obtained in the RAP case and in the plaquette and Polyakov gauges 
in Ref.~\cite{artsu2} is very good.

\subsection{$SU(3)$ gauge theory}

The confinement transition in pure $SU(3)$ gauge theory is a first order
transition~\cite{Fuku}. One therefore expects a pseudocritical
behaviour, with $\nu=1/3$, that is, the inverse of the number of 
spatial dimensions.
As it was remarked in Ref.~\cite{artsu3}, the scaling 
relation, Eq.~(\ref{scaling}), has to be modified in this case to include
finite size violations to scaling,
\begin{equation}
\label{scaling2}
\frac{\rho}{N_s^{1/\nu}} = f\left[ N_s^{1/\nu} 
\left( \beta _C - \beta \right)\right] + \Psi(N_s),
\end{equation}
where $\Psi(N_s)$ parametrizes these effects. A simple assumption is
\begin{equation}
\Psi(N_s)=\frac{a}{N_s^3},
\label{psi1}
\end{equation}
valid up to ${\cal O}(1/N_s^6)$.

\subsubsection{The Random Abelian Projection}

\begin{figure}[tb]
\centerline{\epsfig{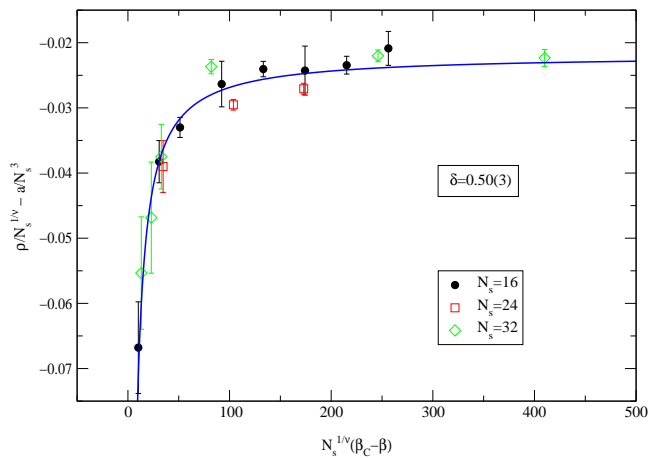}}
\vskip 0.5cm
\caption{Quality of scaling in the RAP case for $SU(3)$.}
\label{fig:su3_nogauge1}
\end{figure}

Fig.~\ref{fig:su3_nogauge1} shows the scaling behaviour expressed
by Eq.~(\ref{scaling2}), where we have taken for $\Psi(N_s)$ the
form~(\ref{psi1}). We used as an input the values $\beta_C(N_t=4)=5.6925$
and $\nu=1/3$~\cite{karschsu3}. The curve in the figure corresponds to a best fit to 
Eq.~(\ref{eqfit}), modified by including 
the term $\Psi(N_s)$, which gives the value $\delta=0.50(3)$,
with a $\chi^2/{\rm d.o.f.}=3.2$. 

\begin{figure}[b]
\centerline{\epsfig{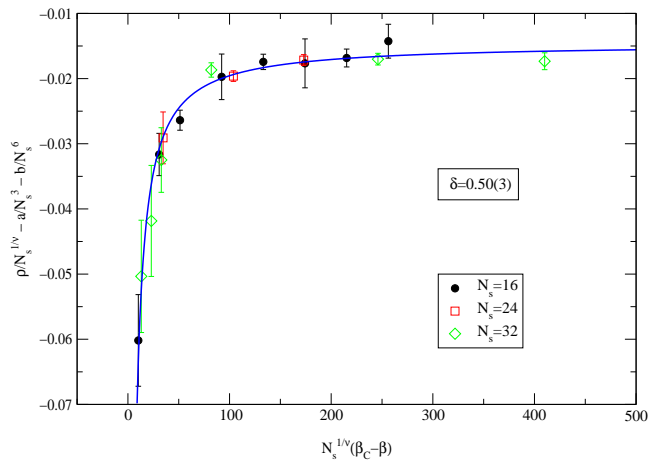}}
\vskip 0.5cm
\caption{The same as in Fig. \ref{fig:su3_nogauge1},
using Eq.~(\ref{psi2}) for $\Psi(N_s)$.}
\label{fig:su3_nogauge2}
\end{figure}

The quality of the fit improves considerably if one includes a second
term in the expression for $\Psi(N_s)$:
\begin{equation}
\Psi(N_s)=\frac{a}{N_s^3} + \frac{b}{N_s^6}.
\label{psi2}
\end{equation}
Then the value of $\delta$ remains the same, while 
$\chi^2/{\rm d.o.f.}=0.7$. 
This demonstrates the importance of the finite size
effects in $SU(3)$ gauge theory. 
The fit is shown in Fig.~\ref{fig:su3_nogauge2}.
\vskip 1.0cm

\subsubsection{The Abelian Projection with Random Ordering}

\begin{figure}[b]
\centerline{\epsfig{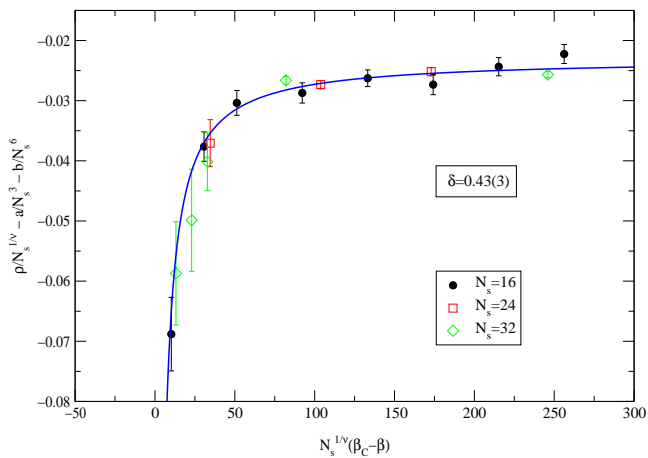}}
\vskip 0.5cm
\caption{Quality of scaling in the APRO case for $SU(3)$,
using Eq.~(\ref{psi2}) for $\Psi(N_s)$.}
\label{fig:su3_noorder2}
\end{figure}

In the APRO case we have obtained a good scaling behaviour as well,
as shown in Fig.~\ref{fig:su3_noorder2}.

However in this case a fit
to Eq.~(\ref{scaling2}) with the function~(\ref{psi1}) has a very bad 
$\chi^2/{\rm d.o.f.}$ (of order 16).

Also in this case the use of the expression~(\ref{psi2}) is essential: 
the best fit, shown in the figure, gives $\delta=0.43(3)$ with 
$\chi^2/{\rm d.o.f.}\sim 1.4$.
The value of $\delta$ is nearly compatible with
the one obtained in the RAP case, $\delta=0.50(3)$. The result obtained
in Ref.~\cite{artsu3} is $\delta=0.54(4)$. 
\vskip 1.0cm

\section{Conclusions}

We have produced further and compelling evidence that monopole condensation
is independent of the Abelian projection used to define the monopoles.

If the idea of duality is correct, the non-local excitations
which are expected to be the fields of the dual description of QCD
and weakly interacting in the confined phase, should have non zero
magnetic charge in all the Abelian projections. This is a very important 
symmetry property, which can help in identifying them.

There has been a number of papers in the literature of the past years,
claiming that the fundamental fields of the dual description 
are the monopoles defined by the maximal Abelian projection.
The claim that monopoles defined by the maximal Abelian projection could be 
the dual excitations does not look in good shape after the quantitative
attempts to construct the dual theory, which go beyond the initial empirical 
observation of Abelian dominance \cite{mondom}. 
If it were true 
maximal Abelian monopoles should be magnetically charged in all 
Abelian projections.
This does not seem to be plausible, since one single Abelian
projection does not confine the $U(1)$ neutral particles belonging to the
adjoint representation,
which would instead be confined in other Abelian projections.

Analysis of the $Z_N$ vortices could give some hints and
investigation has been started in that direction \cite{vortex}.
We think that the problem is still open.

\section*{Acknowledgements}
This work is partially supported by EU contract FMRX-CT97-0122 and
by MURST. 
BL thanks PPARC for a postdoctoral fellowship.

\end{document}